\documentclass{revtex4}

\usepackage{graphicx}
\usepackage{bm}
\usepackage{amssymb}
\usepackage{amsmath}

\begin{document}

\title{Tests of the Envelope Theory in One Dimension}

\author{Claude \surname{Semay}}
\email[E-mail: ]{claude.semay@umons.ac.be}
\author{Lorenzo \surname{Cimino}}
\email[E-mail: ]{lorenzo.cimino@student.umons.ac.be}
\affiliation{Service de Physique Nucl\'{e}aire et Subnucl\'{e}aire,
Universit\'{e} de Mons,
UMONS Research Institute for Complex Systems,
Place du Parc 20, 7000 Mons, Belgium}
\date{\today}

\begin{abstract}
The envelope theory is a simple technique to obtain approximate, but reliable, solutions of many-body systems with identical particles. The accuracy of this method is tested here for two systems in one dimension with pairwise forces. The first one is the fermionic ground state of the analytical Calogero model with linear forces supplemented by inverse-cube forces. The second one is the ground state of up to 100 bosons interacting via a Gaussian potential. Good bounds can be obtained depending on values of the model parameters.
\end{abstract}

\maketitle

\section{Introduction}
\label{sec:intro}

The envelope theory (ET) \cite{hall80,hall83,hall04} is a simple technique to compute approximate solutions, eigenvalues and eigenvectors, of $N$-body systems with arbitrary kinematics in $D$ ($> 2$) dimensions for identical particles \cite{sema13,sema17,sema19}. In the most favourable cases, the approximate eigenvalues are analytical lower or upper bounds. Otherwise, numerical approximation can always be easily computed. The method relies on the existence of an exact solution for the $N$-body harmonic oscillator Hamiltonian, say $\tilde H$. The basic idea is to optimise the eigenvalues of $\tilde H$ to render them as close as possible to those of the Hamiltonian $H$ under study. The accuracy of the method has been checked for various three-dimensional systems containing up to 10 bosons \cite{sema15a,sema15b}. It could be interesting to test the method with an analytical model different from the $N$-body oscillator and with numerical results for a very large number of particles. These two tests can be performed in one dimension space.  

Computations for $D=1$ show that formulas obtained in \cite{sema13,sema19} are still valid, but with the momentums and the positions of particles which are now scalar quantities, and the global quantum number $Q$ with the centre of mass removed which is defined by ($\hbar=1$) 
\begin{equation}
\label{Q}
Q=\sum_{i=1}^{N-1} n_i + \frac{N-1}{2}, 
\end{equation}
where $n_i$ are non-negative integers. For bosons and fermions, the ground state values of $Q$ are respectively \cite{levy68}
\begin{align}
\label{QB}
Q_0^B = \frac{N-1}{2}, \\
\label{QF}
Q_0^F = \frac{N^2-1}{2},
\end{align}
since $Q$ is a global quantum number for a $N$-body harmonic oscillator state. The general Hamiltonian for $N$ identical particles interacting via pairwise forces is given by
\begin{equation}
\label{H}
H=\sum_{i=1}^N T(|p_i|) + \sum_{i < j=2}^N V\left(|x_{ij}|\right),
\end{equation}
where $x_{ij}=x_i-x_j$, with $x_k$ the position of the $k$th particle and $p_k$ the conjugate variable of $x_k$. As only the internal motion is relevant, $\sum_{i=1}^N p_i = 0$. Within the ET, the potential $V(x)$ is in some sense approximated by a potential envelope $\tilde V(x)=c_1\, x^2+c_2$ which is tangent to $V(x)$ for at least one point \cite{hall80,sema13}. The situation is the same for the kinetic part $T(p)$, for which an envelope $\tilde T(p)=d_1\, p^2+d_2$ can also be defined \cite{sema13}. The set of equations giving approximate solutions of this Hamiltonian is given by \cite{sema13,sema19} 
\begin{align}
\label{EAFM1}
E &= N\, T(p_0) + C_N^2 V\left(\frac{x_0}{\sqrt{C_N^2}} \right), \\
\label{EAFM2}
x_0\,p_0 &= Q , \\
\label{EAFM3}
N\, p_0 \,T'\left(p_0 \right)
&= \sqrt{C_N^2}\, x_0\, V' \left(\frac{x_0}{\sqrt{C_N^2}} \right),
\end{align}
where $C_N^2 = \frac{N!}{2!\,(N-2)!}=\frac{N(N-1)}{2}$ and $U'(y)=d U(y)/d y$. The parameter $x_0$ can be determined by (\ref{EAFM3}) taking into account (\ref{EAFM2}). Then, the approximate energy $E$ can be computed with (\ref{EAFM1}). An upper bound is obtained if $\tilde V(x)\ge V(x)\ (\forall\ x\ge 0)$ and $\tilde T(p)\ge T(p)\ (\forall\ p\ge 0)$ \cite{hall83,sema13}. A lower bound is obtained if both ``$\ge$" are replaced by ``$\le$". No bound can be guaranteed in other situations. The accuracy of the method is tested with two one-dimensional systems in the following sections.

\section{Calogero model}
\label{sec:calo}

The Calogero model characterized by a nonrelativistic kinematics, $T(p)=p^2/(2\,m)$, and the pairwise potential, 
\begin{equation}
\label{VCal}
V(x)=\frac{1}{4}\, m\,\omega^2\,x^2+\frac{g}{x^2} 
\end{equation}
with $g>0$, is exactly solvable \cite{calo69}. The ground state energy $E_{\textrm{ex}}$ of $N$ identical fermions is given by
\begin{equation}
\label{ECal}
E_{\textrm{ex}}=\omega\,\sqrt{\frac{N}{2}}\frac{N-1}{2}\left( N+1+N \frac{\sqrt{1+4\,g'}-1}{2} \right),
\end{equation}
where $g'=m\, g$ is dimensionless. 

\begin{figure*}[htb]
\includegraphics[width=0.5\textwidth]{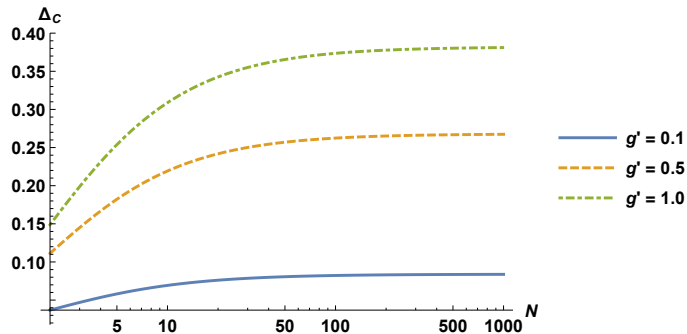}
\caption{Relative error $\Delta_C$ from (\ref{Delta}) as a function of $N$ for three values of $g'$.}
\label{fig:calo} 
\end{figure*}

The ground state lower bound $E_{\textrm{ET}}$ computed with (\ref{EAFM1}-\ref{EAFM3}) and $Q=Q_0^F$ has the following form
\begin{equation}
\label{ECalET}
E_{\textrm{ET}}=\omega\,\sqrt{\frac{N}{2}}\frac{N-1}{2}\sqrt{(N+1)^2+2\,N\,g'}.
\end{equation}
It can be easily checked that $E_{\textrm{ex}}^2 > E_{\textrm{ET}}^2$. As $E_{\textrm{ET}}$ gives the exact result for $g=g'=0$, it can be expected that the lower bound is better for small values of $g'$. This is confirmed on Fig.~\ref{fig:calo} where the relative error 
\begin{equation}
\label{Delta}
\Delta_C= \frac{E_{\textrm{ex}}-E_{\textrm{ET}}}{E_{\textrm{ex}}}
\end{equation}
is plotted as a function of $N$. This error saturates for large values of $N$ at 
\begin{equation}
\label{Deltainf}
\lim_{N\to\infty}\Delta_C= \frac{\sqrt{1+4\,g'}-1}{\sqrt{1+4\,g'}+1}.
\end{equation}
When $g'$ increases, the potential~(\ref{VCal}) deviates more and more from $\tilde V(x)$ at short distance, with a strong and rapid repulsion. This explains the degradation of the lower bound with the increase of $g'$. Nevertheless, the ET can give acceptable results for the Calogero model provided $g' = m\,g < 1$.

\section{Gaussian potential}
\label{sec:gauss}

One of the models computed in \cite{timo17} is characterized by a nonrelativistic kinematics, $T(p)=p^2/(2\,m)$, and the pairwise potential 
\begin{equation}
\label{VGauss}
V(x)= -V_g \, e^{-x^2/a^2}.
\end{equation}
The ground state for a large number of bosons can be accurately computed with an extension of the Lagrange mesh (LM) method \cite{baye15}. Bosonic ground energies $E_{\textrm{LM}}$ are presented in \cite{timo17} for $N=\{3, 5, 20, 100\}$, for the parameters $1/m=43.281307$ (a.u.)$^2$~K, and $V_g = V_0/(\sqrt{\pi}\, a)$ with $V_0=10$~a.u~K and $a=\{0, 0.05, 0.1, 0.2, 0.5, 1\}$~a.u.

\begin{figure*}[htb]
\includegraphics[width=0.48\textwidth]{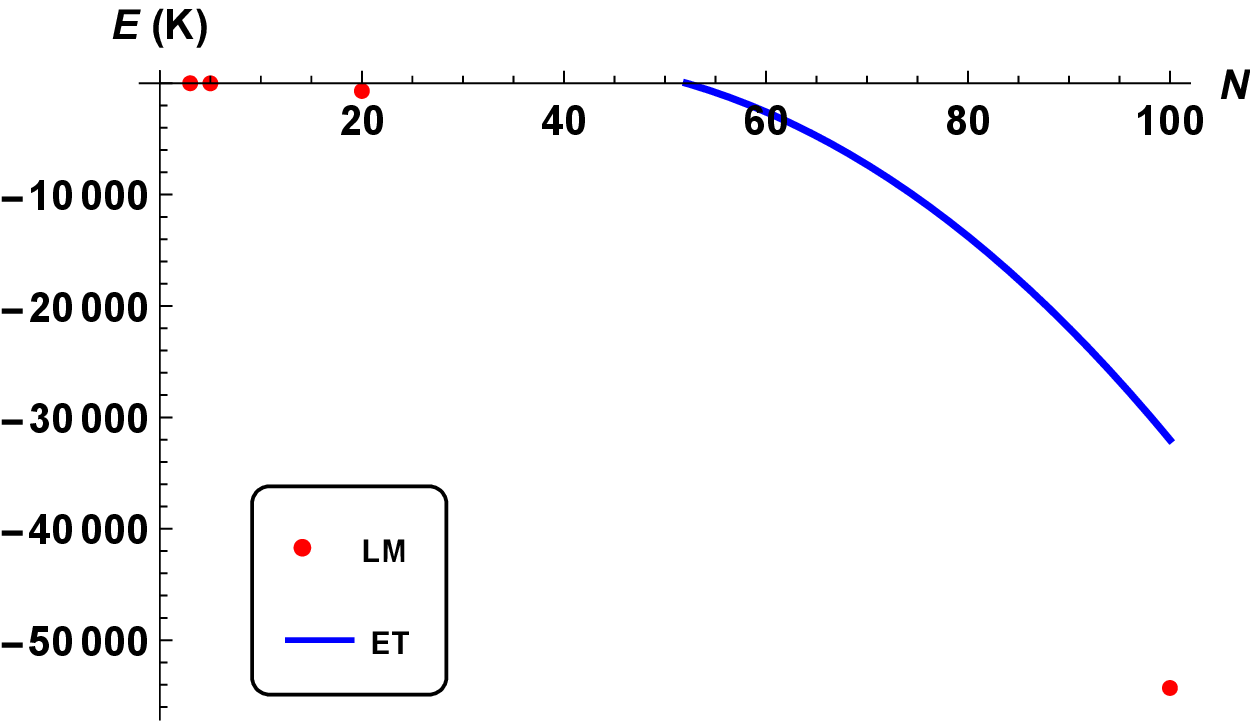} \quad
\includegraphics[width=0.48\textwidth]{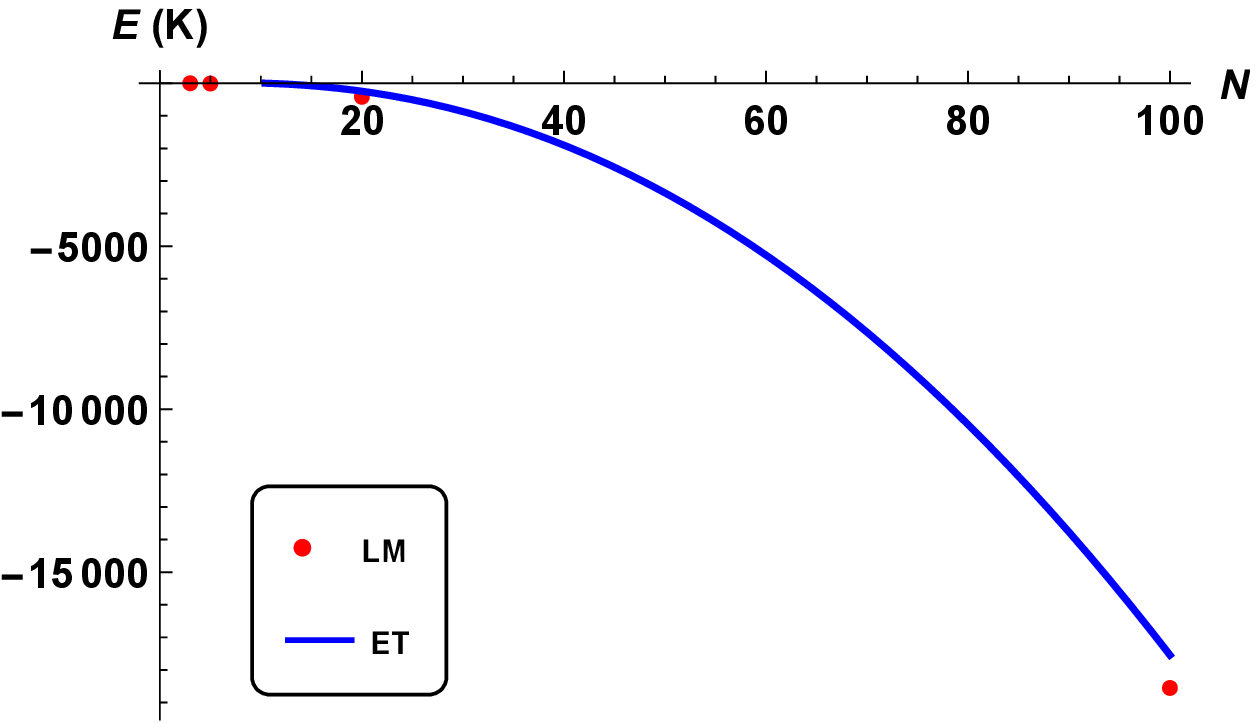} 
\caption{Energy in K for the Gaussian potential as a function of $N$ for $a=0.2$~a.u.\ (left) and $a=1.0$~a.u.\ (right): accurate results of \cite{timo17} obtained with the LM method (dot); upper bounds (\ref{Ewib}) computed with the ET (solid line).}
\label{fig:gauss} 
\end{figure*}

The ground state upper bound $E_{\textrm{ET}}$, computed with (\ref{EAFM1}-\ref{EAFM3}) and $Q=Q_0^B$, is given by \cite{sema15a}
\begin{align}
&E_{\textrm{ET}}=-\frac{N(N-1)}{2}\, V_g\, Y^2\, \frac{1+2\, W_0(Y)}{W_0(Y)^2} \nonumber\\
\label{Ewib}
&\textrm{with}\quad Y=-\frac{1}{2\, a \sqrt{2\,m\, V_g\, N}}.
\end{align}
The multivalued Lambert function $W(z)$ is the inverse function of $z\,e^z$ \cite{corl96}. $W_0(z)$ is the branch defined for $z \ge -1/e$. The quality of this upper bound can be checked on Fig.~\ref{fig:gauss} for two values of $a$. 

Let us first remark that irrelevant values for the energies (complex or positive real numbers) are computed with the ET when the number of particles is too small. Such a behaviour was already observed for $D=3$ \cite{sema15a}. The quality of the bound strongly depends on the value of $a$. When this range is small, the potential has a large variation ($V(0)\propto a^{-1}$) on the short distance $a$, and it is approximated with more difficulty by the potential envelope $\tilde V(x)$. Table~\ref{tab:1} shows that the relative error
\begin{equation}
\label{Delta2}
\Delta_G= \frac{E_{\textrm{LM}}-E_{\textrm{ET}}}{E_{\textrm{LM}}}
\end{equation}
decreases when $N$ or $a$ increase. If the potential does not vary too rapidly, a reasonable accuracy can be reached for a large number of particles, where the numerical computations can be lengthy. 

\begin{table}[htp]
\caption{Relative error $\Delta_G$ from (\ref{Delta2}) for several values of $N$ and $a$. No relevant value of the upper bound can be computed for $N=20$ and $a=0.2$~a.u.}
\begin{center}
\label{tab:1}
\begin{tabular}{cccc}
\hline\noalign{\smallskip}
 & \multicolumn{3}{c}{$a$ (a.u.)} \\
$N$ & 0.2 & 0.5 & 1.0 \\
\noalign{\smallskip}\hline\noalign{\smallskip}
20 & - & 1.06 & 0.41 \\
100 & 0.41 & 0.13 & 0.054 \\
\noalign{\smallskip}\hline
\end{tabular}
\end{center}
\end{table}

\section{Concluding remarks}
\label{sec:conclu}

The ET has already been tested in the $D=3$ space for various Hamiltonians up to 10 bosons \cite{sema15a,sema15b}. It was shown that reliable results can be obtained for energies and some observables. The method is tested here in the $D=1$ space with two systems: the fermionic ground state of a Calogero model and the ground state of up to 100 bosons interacting via a Gaussian potential. It is clear with the examples presented that the quality of the bound greatly depends on the parameters of the system. The problem is that the parameters allowing good approximations cannot be systematically predicted. Nevertheless, the ET is so simple to implement that it is worth using it. If a great accuracy is not searched for, the approximations supplied can be sufficient. Otherwise, the approximations computed can be used as tests for accurate numerical computations.     

The ET is tested here in favourable situations where a bound can be computed analytically. If it is not the case, numerical approximations, with or without variational character, can always be computed. At the price of loosing the possible variational character of the ET, the approximate energies can be improved by modifying the structure of the global quantum number $Q$ \cite{sema15b}. But this is possible only for states in $D \ge 2$ spaces, in which an angular momentum can be defined. This is not the case for the $D=1$ space.

\end{document}